# Dielectric spectroscopy on aging glasses

P. Lunkenheimer[*], R. Wehn, A. Loidl

*Experimental Physics V, Center for Electronic Correlations and Magnetism, University of Augsburg, 86135 Augsburg, Germany*

**Abstract**

In the present work, we provide further evidence for the applicability of a modified stretched-exponential behavior, proposed recently for the description of aging-time dependent data below the glass temperature [P. Lunkenheimer *et al.*, Phys. Rev. Lett. 95 (2005) 055702]. We analyze time-dependent dielectric loss data in a variety of aging glasses, including new data on Salol and propylene carbonate, using a conventional stretched exponential and the newly proposed approach. Also the scaling of aging data obtained at different measuring frequencies, which was predicted on the basis of the new approach, is checked for its validity.

## 1. Introduction

The dynamics of glassy matter exhibits some common phenomena, which can be considered as hallmark features of glassy behavior, namely stretching, non-Arrhenius behavior, and ergodicity breaking [1]. "Stretching" denotes the non-exponential relaxational response of glassy matter to external perturbations, which nowadays is commonly ascribed to heterogeneity leading to a distribution of relaxation times. Non-Arrhenius behavior shows up in the temperature dependence of dynamic quantities as, e.g., the viscosity or the relaxation time and is often assumed to mirror increasingly cooperative molecular motions when approaching the glass transition. While those phenomena usually are observed at temperatures above the glass temperature $T_g$, ergodicity breaking arises when the sample "falls out of equilibrium" under cooling, which for typical cooling rates occurs close to $T_g$. This leaves the sample in a structural state corresponding to equilibrium at a higher temperature and leads to the observation of a variety of different, partly quite intriguing phenomena as aging, memory effects, and rejuvenation. From a theoretical viewpoint, ergodicity breaking and the resulting non-equilibrium processes are among the most challenging phenomena of glassy dynamics. They are also of considerable practical interest, e.g. for polymers, usually applied at temperatures not too far below $T_g$, where aging can lead to degradations of material properties. A famous approach for the description of non-equilibrium effects in glasses is the Tool-Narayanaswamy-Moynihan (TNM) formalism [2,3,4]. By introducing the concepts of fictive temperature and reduced time, it takes into account the so-called non-linearity of structural relaxation during aging, caused by the fact that the relaxation time itself is time-dependent.

Aging denotes the time-dependent variation of physical quantities when a glass-forming sample reapproaches equilibrium after quenching it below $T_g$. This phenomenon was termed "physical aging" by Struick [5], to distinguish it from time-dependent processes involving chemical reactions; alternative expressions used in literature are "annealing" (if the time-dependent changes are intentional) and "structural relaxation". Monitoring the time-dependence of physical quantities during physical aging is a straightforward experiment for the investigation of non-equilibrium effects. While there are numerous recent reports on physical aging in spin glasses [6] and polymers [5,7,8,9], much fewer experiments were performed on canonical glass formers. Especially, having in mind that dielectric spectroscopy has proven a key technique for the study of glassy dynamics in equilibrium, it is astonishing that investigations of glassy non-equilibrium dynamics in low molecular-weight glass formers with this technique are relatively scarce [3,10,11,12].

---

[*] Corresponding author. Tel. +49 821 598 3603; fax: +49 821 598 3649; E-mail: peter.lunkenheimer@physik.uni-augsburg.de

Thus, in a recent work [13], we have provided detailed dielectric aging data on a variety of materials, belonging to different classes of glass formers, namely molecular glass formers and a glass-forming ionic melt, having different fragilities [1,14,15], bonding types, and being characterized by an excess wing [16] or a well-developed Johari-Goldstein β-relaxation [17]. In that work we have introduced a new type of modeling of the experimental data by using a self-consistent recursive formula describing the time-dependence of physical properties of the glass. Using this approach we could demonstrate that the aging dynamics in all these materials is fully determined by the relaxation time and stretching parameter of the α-relaxation. In the present work, we provide further evidence for the validity of this approach by applying it to further dielectric aging data and by comparing it with an evaluation using a conventional stretched exponential ansatz. In addition, we check the data for the scaling properties, predicted on the basis of the new approach.

## 2. Experimental procedures

For the measurements, parallel plane capacitors having an empty capacitance up 100 pF were used. High-precision measurements of the dielectric permittivity in the frequency range $10^{-4} \leq n \leq 10^6$ Hz were performed using a frequency response analyzer. At selected temperatures and aging times, additional frequency sweeps at $20\,\text{Hz} \leq n \leq 10^6\,\text{Hz}$ were performed with an autobalance bridge [18]. To keep the samples at a fixed temperature for up to five weeks, a closed-cycle refrigerator system was used. The samples were cooled from a temperature at least 20 K above $T_g$ with the maximum possible cooling rate of about 3 K/min. The final temperature was reached without any temperature undershoot. As zero point of the aging times $t_{age}$, we took the time when the desired temperature was reached, typically about 100 s after passing $T_g$. The temperature was kept stable better than 0.1 K for all aging measurements.

## 3. Results

Figure 1 shows the time dependence of the dielectric loss $e''$ of four molecular glass formers, namely propylene carbonate (PC, $T_g \approx 159$ K, $m = 104$), propylene glycol (PG, $T_g \approx 168$ K, $m = 52$), glycerol $T_g \approx 185$ K, $m = 53$), and Salol ($T_g \approx 218$ K, $m = 73$), after quenching them to a temperature several K below $T_g$. Here $m$, taken from [15], denotes the so-called fragility index [19] giving a measure of the departure of the temperature-dependent relaxation time from Arrhenius-behavior (the lower limit of the fragility parameter is $m \approx 16$, the most fragile materials have $m \leq 200$). As revealed by Fig. 1, $e''$ of all materials continuously decreases during aging. For long aging times $t_{age}$, the loss of PC and glycerol, which have been aged up to five weeks, finally becomes constant and for PG and Salol it at least shows the tendency to saturate. This indicates that the thermodynamic equilibrium state is reached after about $10^6$ s. The lines are fits with a stretched exponential law as described in detail in section 3.

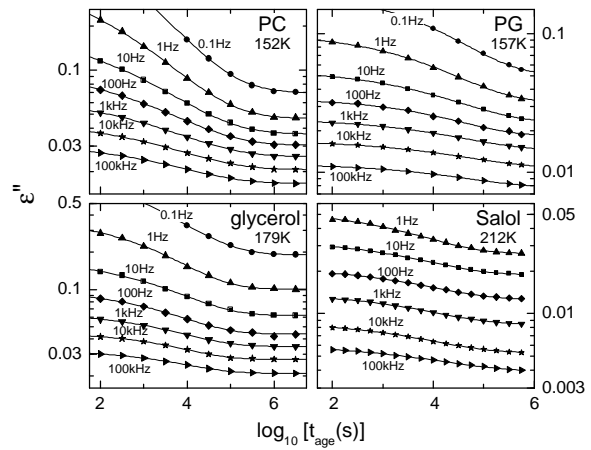

Fig. 1. Aging-time dependence of the dielectric loss of four different glass formers for various frequencies (the data on PC, PG, and glycerol were taken from [13]). The lines are fits with eq. (1). The errors are of similar magnitude as the width of the symbols.

In Fig. 2(a), we present aging data for PC collected at an additional lower temperature. These data and those on Salol at 212K [Fig. 2(b)] are fitted with the modified KWW law, introduced in [13] as described in the discussion section.

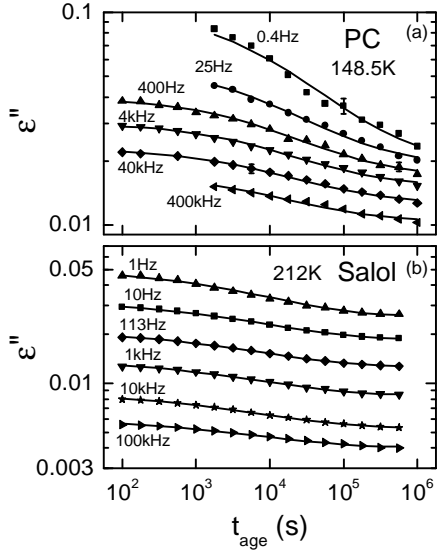

Fig. 2 Aging-time dependence of the dielectric loss of PC and Salol for various frequencies, fitted with the new approach, eq. (2), introduced in [13]. The values of $b_{age}$ were fixed to $b_{age} = 0.6$ (PC [16,26]) and 0.4 (Salol [27]). For PC, in addition the equilibrium relaxation time was fixed to $t_{eq} = 1.6 \times 10^8$ s. The errors are of similar magnitude as the width of the symbols.

## 4. Discussion

The solid lines in Fig. 1 [20,21] show fits of these aging data using the time-honored Kohlrausch-Williams-Watts (KWW) law, routinely applied to describe *equilibrium* relaxation processes in supercooled liquids [1]:

$$e''(t_{age}) = (e''_{st} - e''_{eq}) \exp\left[-(t_{age}/t_{age})^{b_{age}}\right] + e''_{eq} \quad (1)$$

Here the indices "st" and "eq" indicate the values for $t_{age} \to 0$ and $\infty$, respectively (both being fitting parameters), $t_{age}$ represents the relaxation time and $b_{age}$ the stretching parameter [12,22]. Fits and experimental data match perfectly using this ansatz (Fig. 1).

As shown in Fig. 3 the resulting fit parameters $t_{age}$ and $b_{age}$ exhibit a considerable frequency dependence. Such a behavior was also found in the dielectric study of aging in glycerol by Leheny and Nagel [12]. A frequency-dependence of these parameters may be rationalized only if for different frequencies, different dynamic processes prevail. In the present cases, by taking into account the equilibrium spectra obtained close to $T_g$ [16,23,24], it can be deduced that at the lowest frequencies indeed the aging of the high-frequency flank of the α-process is detected while at the higher frequencies the excess wing or the corresponding secondary relaxation process [16,23] governs the response. Thus one could explain the observed frequency dependence in Fig. 3 by assuming that the α-relaxation and the secondary relaxation may be governed by different aging dynamics.

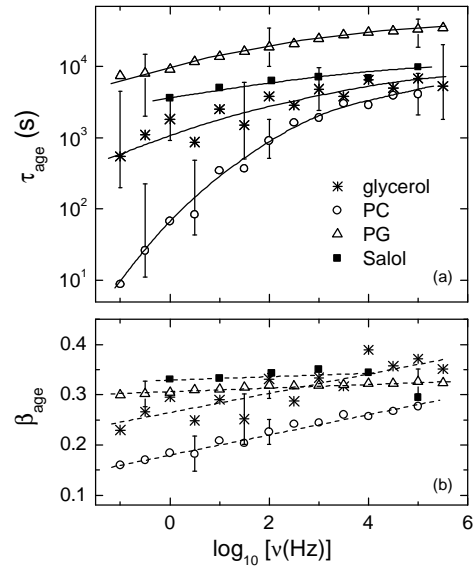

Fig. 3. Parameters of the fits with eq. (1) to $e''(t_{age})$ shown in Fig. 1.

However, there is another severe problem with the results of Fig. 3: Neither the resulting $t_{age}$ nor $b_{age}$ agree with the corresponding parameters $t_\alpha$ and $b_\alpha$, determined from an extrapolation of the equilibrium values obtained at $T > T_g$. Especially, fits with eq. (1) yield $b_{age}$ always much smaller than the equilibrium value [12,25]. At first glance, this seems difficult to understand as it is reasonable that the variation of $e''$ during aging can be traced back to structural rearrangements and thus should be governed by the same dynamics (i.e. relaxation time) and heterogeneity (i.e. stretching parameter) as the α-relaxation. For the materials of Fig. 1, the extrapolated $b_\alpha$ are 0.6 (PC) [26], 0.58 (PG) [24], 0.55 (glycerol) [26], and 0.4 (Salol) [27]. These values all are significantly larger than the fit results from eq. (1), shown in Fig. 3(b) and it is not possible to fit the data of Fig. 1 with $b_{age}$ fixed to the extrapolated value [28]. Similar discrepancies also show up if comparing the extrapolated $t_a$ values with those shown in Fig. 3(a) [13].

These discrepancies are the manifestation of the so-called non-linearity of sub-$T_g$ relaxation, a phenomenon addressed already in the famous works by Tool and Narayanaswamy [2]. Namely it was pointed out that during aging, $t_{age}$ itself is time dependent. In the Tool-Naranayaswamy-Moynihan (TNM) formalism [2,3,4] this is taken into account by tracing back the aging-induced variation of physical quantities to the time-dependence of the so-called fictive temperature $T_f$ and by introducing an additional non-linearity parameter. This formalism (and other, mathematically nearly equivalent ones [11,29]) has been successfully used to describe various aging experiments (see, e.g., [4]). However, its application is not straightforward, involving the numerical solution of a set of equations and requiring some assumptions for a proper evaluation. Instead, in [13] we have proposed a new approach that is much simpler to apply. Considering that the α-peak frequency $n_p$ is a good estimate of the relaxation rate $n_\alpha = 1/(2\pi t_\alpha)$, the fact that during aging the α-relaxation peak shifts to lower frequencies [13] reflects the decrease of the relaxation rate due to the gradual drop of $T_f$ towards the actual temperature. Thus, obviously the relaxation rate itself is subjected to aging. To take into account this time dependence, we make an ansatz for $n_\alpha(t_{age})$, analogous to eq. (1), namely:

$$n_a(t_{age}) = 1/(2\pi t_a) = \\ = (n_{st} - n_{eq}) \exp\left[-\left(t_{age} 2\pi n_a\right)^{b_a}\right] + n_{eq} \quad (2)$$

One should note here that the time scale determining the aging of the relaxation rate is given by the inverse relaxation rate (i.e. the relaxation time) itself. This ansatz implies that the stretching (quantified by $b_\alpha$) remains unaffected by aging, i.e. that time-temperature superposition is valid (this was evidenced, e.g., for a polymeric system in [9]). As the shift of $T_f$ during aging is only few K for these experiments, this assumption is justified.

Equation (2) can be easily solved numerically by recursion [30], resulting in an age dependence of $t_a$ that takes into account the non-linearity effects discussed above. Assuming that $t_{age} = t_a$ and $b_{age} = b_a$, the obtained $t_{age}(t_{age})$ is put into eq. (1), which then is used to fit the measured $e''(t_{age})$. In [13] we have demonstrated that the aging data obtained on five different glass formers, including PC, PG, and glycerol could be well described using this approach. The parameters were kept identical for all frequencies and $b_{age}$ was fixed to the extrapolated equilibrium value. The resulting $t_{eq}$ values perfectly match the extrapolated equilibrium $t_\alpha(T)$ curves.

In Fig. 2, we give further evidence for the applicability of the modified KWW ansatz by applying it to aging data on Salol and PC, the latter having been measured at a lower temperature than the data presented in [13]. The lines in Fig. 2 demonstrate the very good fits that are possible using this ansatz. We want to emphasize that the fits were performed simultaneously for *all* curves, with the parameters of eq. (2) identical for *all* frequencies. The curves for different frequencies are only distinguished by the parameters $e''_{st}$ and $e''_{eq}$. Thus the number of parameters for each curve is four ($b$ being fixed to the extrapolated equilibrium value), however, two of them being common for all curves. Thus much less parameters are involved, compared to the conventional KWW fits of Fig. 1. From the fits, we obtain an average relaxation time $<t_{eq}> = 3.7\times10^4$ s for Salol, which matches perfectly the published $<t>(T)$ behavior of Salol [27,31], extrapolated to 212 K. For PC at 148.5 K [Fig. 2(a)] the problem arises that this temperature is too far below $T_g$ to come close to thermodynamic equilibrium within reasonable times, especially as this is a rather fragile glass former with a very steep $t(T)$ curve close to the glass transition. Thus from the fits shown in Fig. 2, it is not possible to gather reliable information on the equilibrium relaxation time. Therefore $n_{eq}$ was fixed to a value of $1\times10^{-9}$ Hz, corresponding to $<t_{eq}> = 2.4\times10^4$ s, consistent with an extrapolation of $t(T)$ of PC [26,32]. Thus, also for this lower temperature, in PC the aging results can be described in full accord with the equilibrium results on the α-relaxation, using the new approach proposed in [13].

The fact that the fits for different frequencies in Fig. 2 only differ by the values of $e''_{st}$ and $e''_{eq}$ implies that it should be possible to scale the $e''(t_{age})$ curves for different frequencies onto one master curve by plotting the quantity $(e'' - e''_{eq}) / (e''_{st} - e''_{eq})$ vs. the aging time. In [13] a nearly perfect scaling was reported for glycerol. In Fig. 4 we provide the scaling plots for two of the glass formers analyzed in [13] and for the new data on PC and Salol of the present work. In all cases the scaling works reasonably well. The lines show the fit curves, which naturally collapse onto one line in the scaling plots. For PC the scaling is least perfect and for this material the scaling plot also reveals systematic deviations of fit and experimental data at the longest aging times. Probably this is due to the fact that for this rather low temperature, the sample is still far from equilibrium and the fixed value of $t_{eq}$ should be chosen even larger. However, it is also possible that for larger temperature jumps limitations of the introduced modified KWW approach show up. In this context it is interesting that in [12], for temperatures far below $T_g$, deviations from the TNM model were reported.

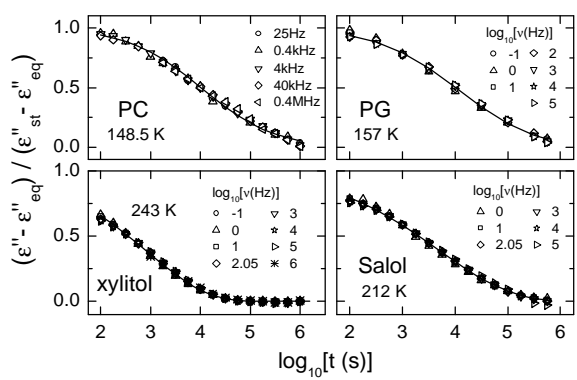

Fig. 4. Scaling of the aging curves for different frequencies for four glass formers. The values of $e''_{st}$ and $e''_{eq}$ were obtained from the fits with the modified KWW approach, shown in Fig. 2 (PC and Salol) and in [13] (PG and xylitol). For PC the data set at 0.4 Hz was omitted as it shows rather strong scattering [cf. Fig. 2(a)]. The lines correspond to the fitting curves.

## 5. Conclusions

In conclusion, the analysis of aging data on a number of different glass formers using a conventional KWW ansatz, performed in the present work, has revealed marked deviations of the resulting relaxation parameters from those obtained from equilibrium data. In contrast, the alternative approach proposed in [13], which takes into account the time dependence of the relaxation time, successfully describes results at different frequencies with identical relaxation time and stretching parameter, both being fully consistent with equilibrium data. In the present work, this approach was successfully applied to time-dependent data on Salol and PC, giving further evidence for its applicability for the description of aging data. Our results also corroborate the conclusion drawn in [13] that all dynamic processes age in a similar way, determined by the structural α-relaxation dynamics.

## Acknowledgements

We thank C.A. Angell, R. Böhmer, R.V. Chamberlin, J.C. Dyre, G.P. Johari, N.B. Olsen, and R. Richert for illuminating discussions. We thank U. Schneider for performing part of the measurements.